# Les incompatibilités entre la théorie standard de l'extinction interstellaire et l'observation

Note soumise à l'Académie des Sciences (Février 2002)

Frédéric Zagury   (fzagury@wanadoo.fr)
02210 Saint Rémy Blanzy, France

Résumé:

L'interprétation standard de la courbe d'extinction de la lumière par la matière interstellaire n'est pas en accord avec certaines observations. Cela s'explique seulement si la lumière des étoiles rougies est contaminée par de la lumière diffusée à très faible angle de diffusion. La courbe d'extinction exacte est alors une ligne droite du proche infrarouge jusqu'à l'UV lointain. Cela remet en question tous les modèles de poussière utilisés à l'heure actuelle. Il n'y a également pas lieu de supposer de variation des propriétés moyennes des grains d'une région de l'espace à l'autre.

Abstract:

The standard interpretation of the extinction curve does not agree with some observations. The light we receive from a reddened star must be contaminated by starlight scattered at very small angular distances from the star. The true extinction curve is a straight line from the near infrared to the far-UV. If so, all interstellar dust models must be questionned. A second conclusion is that the average properties of interstellar grains may be much more uniform than previously thought.

**Extinction / poussière interstellaire**

# 1. Introduction

La poussière interstellaire représente une très faible partie de la matière interstellaire, essentiellement gazeuse. Pourtant, en éteignant la lumière des étoiles, en participant à la diffusion de la lumière stellaire dans les nébuleuses, en étant responsable de l'émission infra-rouge des nuages, la poussière interstellaire joue un rôle considérable comme traceur de la matière interstellaire et de son organisation. Elle est aussi impliquée dans la chimie interstellaire. L'étude de la poussière interstellaire a donné lieu à différents modèles de poussières, dont la principale contrainte est la courbe d'extinction   déduite de l'observation d'étoiles à travers les nuages interstellaires. On conçoit l'importance d'une interprétation juste de cette courbe pour notre compréhension du milieu interstellaire.

L'interprétation admise depuis une trentaine d'années -lorsque les satellites ont permis d'obtenir les premiers spectres UV des étoiles- de la courbe d'extinction de la lumière par la poussière interstellaire sépare, entre le proche infra-rouge et l'UV lointain, trois domaines spectraux: le visible, la région du 'bump' à 2200 Å, et l'UV lointain. Dans chacun de ces domaines un type et une distribution de taille de particules interstellaires sont supposé être responsables de l'extinction. Pour expliquer les larges variations de la courbe d'extinction d'une région à l'autre les proportions relatives de chaque type de grains doivent dépendre de la région considérée. Tous les modèles de poussière en cours à l'heure actuelle sont basés sur ces consensus.

Des observations simples montrent que cette interprétation standard de la courbe d'extinction ne tient pas. Les conséquences sont importantes, car il faut re-considérer la nature de la lumière qui nous parvient d'une étoile lorsque de la matière s'interpose entre l'étoile et l'observateur. L'interprétation des observations de la matière interstellaire, la façon d'analyser ces observations s'en trouvent profondément modifiées; cela remet en question tous les modèles de poussière utilisés actuellement. Les applications sont multiples, allant de problèmes pratiques comme la correction du rougissement pour estimer la distance des étoiles, à des problèmes plus théoriques sur la nature de la poussière interstellaire, ses éventuels modifications selon l'environnement, etc..

Je discuterai successivement l'interprétation standard de la courbe d'extinction, les observations qui permettent d'en douter, l'interprétation de ces observations et quelques conséquences.

## 2. La courbe d'extinction et son interprétation classique

Une étoile dont la lumière à la longueur d'onde $\lambda$ traverse un nuage d'épaisseur optique $\tau_\lambda$ voit son flux atténué d'un facteur $e^{-\tau_\lambda}$. Si $F_\lambda$ et $F_{0\lambda}$ sont le flux mesuré et celui de l'étoile si elle n'était pas rougie:

$F_\lambda = F_{0\lambda}\, e^{-\tau_\lambda}$

En magnitudes:

$m_\lambda = m_{0\lambda} + 1.07\, \tau_\lambda = m_{0\lambda} + A_\lambda$

$F_{0\lambda}$ et $m_{0\lambda}$ n'étant pas connus sont remplacés par les valeurs pour une étoile non rougie de même type spectral. Cela permet d'obtenir la courbe d'extinction $A_\lambda$ à une constante près, généralement déterminée à partir des magnitudes V des étoiles. Cette courbe est indépendante du spectre propre de l'étoile; elle caractérise la colonne densité de matière entre l'étoile et nous. La courbe d'extinction peut être normalisé par $E(B-V) = A_B - A_V$ (proportionnel à la pente de l'extinction linéaire observée dans le visible) pour tenir compte des différences de colonne densité de matière entre les différentes directions de l'espace.

Seaton [1] a donné la courbe d'extinction normalisée moyenne pour les étoiles du voisinage solaire (figure 1). La théorie standard sépare cette courbe en trois parties: le visible, la région du bump à 2200 Å, et l'UV lointain. La courbe d'extinction normalisée en direction d'une étoile rougie suit la courbe de Seaton sur la partie visible et linéaire, mais de fortes variations d'une étoile à l'autre existent au niveau du bump et surtout dans l'UV lointain [2, 3, 4]. Dans la théorie standard ces variations sont expliquées par la présence de trois types de particules présents en proportions variables dans les différents nuages (voir la figure 1). L'extinction linéaire en $1/\lambda$ du visible est attribuée à une distribution de "gros" grains. La région du "bump" est attribuée à de l'absorption par des petites particules. Enfin la croissance de la courbe d'extinction dans l'UV lointain serait due à de l'extinction par des molécules, e.g. poly-aromatiques (PAH). Les qualificatifs "gros" ou "petits" s'entendent en comparaison avec le domaine de longueur d'onde considéré, le rapport entre la dimension du grain et la longueur d'onde étant un paramètre essentiel de la théorie de l'extinction.

En faisant varier la proportion de chaque type de particules, la théorie standard se donne trois degrés de liberté qui permettent de reproduire les courbes d'extinction.

Cette liberté que se donne la théorie standard est contre-balancée par deux contreparties non négligeables.

D'abord, il ne semble pas y avoir de cohérence entre les variations de la courbe de région en région et l'environnement: aucune des recherches visant à trouver une logique dans la répartition des trois types de grain suivant l'environnement (densité, exposition aux UV…) n'est véritablement concluante [5].

Ensuite, chaque type de particules de la théorie standard a une courbe d'extinction (figure 1) suffisemment particulière pour contraindre fortement la distribution en taille et la nature des particules.

En pratique, aucun des modèles en cours à l'heure actuelle (le "modèle unifié" de Greenberg et Li [6, 7], le modèle de Mathis [8], celui proposé par Désert et al. [9], etc..) n'est complètement satisfaisant.

## 3. La théorie standard et les observations

Différentes observations permettent de tester de façon fine les hypothèses de base de la théorie standard. J'ai proposé quatre tests, dont trois sont explicités ci-après.

### 3.1 Le spectre UV des nébuleuses

J'ai établi [10] que le spectre UV d'une nébuleuse est bien reproduit par le produit du spectre de l'étoile qui l'éclaire et d'une fonction linéaire de $1/\lambda$. Ce rapport linéaire en $1/\lambda$ qui existe entre le spectre d'une nébuleuse et celui de l'étoile qui l'éclaire suggère que la loi de diffusion dans le visible (extinction en $1/\lambda$ et albedo constante) se prolonge dans l'UV.

Ni les gros grains censés faire l'extinction visible, dont l'extinction dans l'UV lointain est quasi indépendante de la longueur d'onde (figure 1), ni les petites particules censées faire l'extinction dans l'UV, qui à priori devraient diffuser en $1/\lambda^4$, ne peuvent expliquer une diffusion linéaire en $1/\lambda$.

Il n'y a pas d'excès de diffusion dans la région du bump, ce qui montre que si des particules éteignent la lumière à 2200Å, elles doivent être absorbantes. Pourtant, certaines nébuleuses, associées à des étoiles faiblement rougies, n'ont pas de bump [10]. Faut-il en conclure que les petits grains responsables du bump à 2200Å n'existent pas dans les milieux diffus? Ces remarques rendent difficile l'explication standard du bump par un processus d'extinction classique.

### 3.2 La courbe d'extinction dans les directions de très faible rougissement (E(B-V)<0.1)

Les spectres d'étoiles de même type spectral et de très faible rougissement, rougies mais pas suffisamment pour présenter un bump (il faut que E(B-V) soit inférieur à env. 0.1) se déduisent l'un de l'autre par une même exponentielle de $1/\lambda$ dans le visible et dans l'UV [11]. Ainsi, pour de très faibles colonnes densité de matière, la loi d'extinction est la même dans le visible et dans l'UV, confirmant ce que l'observation des nébuleuses laisse pressentir.

### 3.3 La courbe d'extinction dans les directions de rougissement faible (0.1≤E(B-V)≤0.4)

Si on augmente graduellement le rougissement, pour des étoiles de rougissement faible bien que plus important que précédemment (0.1≤E(B-V)≤0.4), l'extinction linéaire du visible se prolonge dans le proche UV jusque dans la région du bump [12], ce que ne peut expliquer la théorie standard. Le comportement particulier de la courbe d'extinction dans l'UV lointain se présente nettement comme une composante additionnelle qui se greffe sur la queue de cette extinction (figure 2).

## 4. Discussion

Ces trois types d'observations contredisent de façon manifeste la théorie standard de l'extinction, et n'offrent qu'une alternative [2, 12]: le spectre des étoiles rougies doit être contaminé par de la lumière diffusée. Toute observation d'une étoile rougie comprend une part plus ou moins importante, selon la région traversée par la lumière et selon la longueur d'onde, mais non négligeable, de lumière diffusée. Cette hypothèse avait été rejetée par Snow et York [13], qui n'ont pas observé de différence en analysant le spectre UV de σ-Sco, observé avec deux ouvertures différentes, de 8'x3° et de 0.3"x39". En fait, les observations de Snow et York démontrent seulement que si de la lumière diffusée contamine le spectre des étoiles rougies, cela doit se faire à faible distance angulaire de l'étoile, inférieure à 0.3".

Les observations de la section 3 s'expliquent si la lumière reçue d'une étoile rougie comporte une proportion non négligeable de lumière diffusée. Lorsque la colonne densité de matière entre l'étoile et l'observateur est très faible, la lumière diffusée est négligeable devant celle qui arrive directement de l'étoile; on observe alors la lumière directe de l'étoile seulement, faiblement éteinte, et l'extinction observée reflète la loi d'extinction de la lumière par la poussière interstellaire. Lorsque la colonne densité augmente, c'est d'abord dans l'UV lointain, là où l'extinction est la plus forte, que l'effet de la lumière diffusée est important, puisque l'extinction, et donc le nombre de photons susceptibles d'être diffusés, augmente comme l'inverse de la longueur d'onde. La courbe d'extinction observée dans l'UV diverge alors du prolongement de l'extinction linéaire du visible. Si le rougissement est encore augmenté, la lumière diffusée va également contaminer la partie visible du spectre, provoquant l'écart de la courbe de Seaton à l'extinction linéaire que l'on remarque entre $1/\lambda \sim 2.5 \mu m^{-1}$ et $1/\lambda \sim 4 \mu m^{-1}$ (figure 1).

Les conséquences sont nombreuses. La première conséquence est que la courbe d'extinction est une ligne droite du proche infra-rouge jusque dans l'UV lointain. La seconde conséquence est qu'il n'y a a priori pas de raison de supposer de différence des propriétés moyennes de la poussière interstellaire d'une direction de l'espace à l'autre. La diffusion de la lumière d'une étoile par une nébuleuse, par exemple, suit la même loi en $1/\lambda$ pour toutes les nébuleuses. Si dans un proche avenir l'obtention de spectres de nébuleuses confirme que cette loi est aussi observée dans le visible, nous pourrons également conclure que l'albedo et la fonction de phase des grains sont indépendants de la longueur d'onde. Les variations observées en direction de certaines étoiles du paramètre Rv=Av/E(B-V), souvent utilisées comme preuve de variation des propriétés des grains, doivent être dues à la modification du spectre visible de l'étoile par la lumière diffusée.

Il s'ensuit naturellement que tous les modèles en cours concernant la poussière interstellaire, basés sur une interprétation trop directe des particularités de la courbe d'extinction, sont remis en question.

Ce travail montre parallèlement la nécessité d'obtenir des spectres d'étoiles rougies sur une gamme de longueur d'onde la plus large possible, et d'éviter de travailler avec le logarithme des spectres. Associer simultanément l'étude visible et UV devient nécessaire pour comprendre et séparer les contributions de la lumière directe de l'étoile et celle de la lumière diffusée. Le passage au logarithme du spectre, comme cela se fait habituellement, qui facilite l'étude de l'épaisseur optique lorsqu'il n'y a pas de lumière diffusée, fait qu'il devient difficile de reconnaître et de séparer les deux composantes (additives) qui le composent.

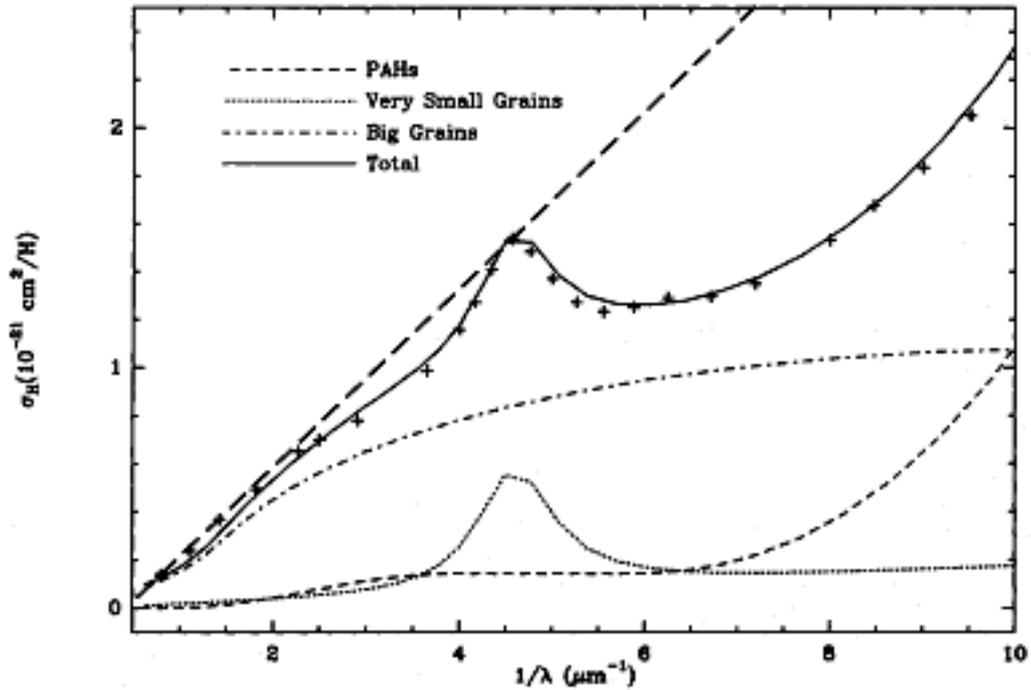

Figure 1 : La courbe d'extinction de Seaton (trait plein) et les contributions séparées des gros grains, des très petits grains (VSG) et des molécules poly-aomatiques-hydrogénées. D'aprés [9]. J'ai ajouté en tirets la courbe d'extinction linéaire.

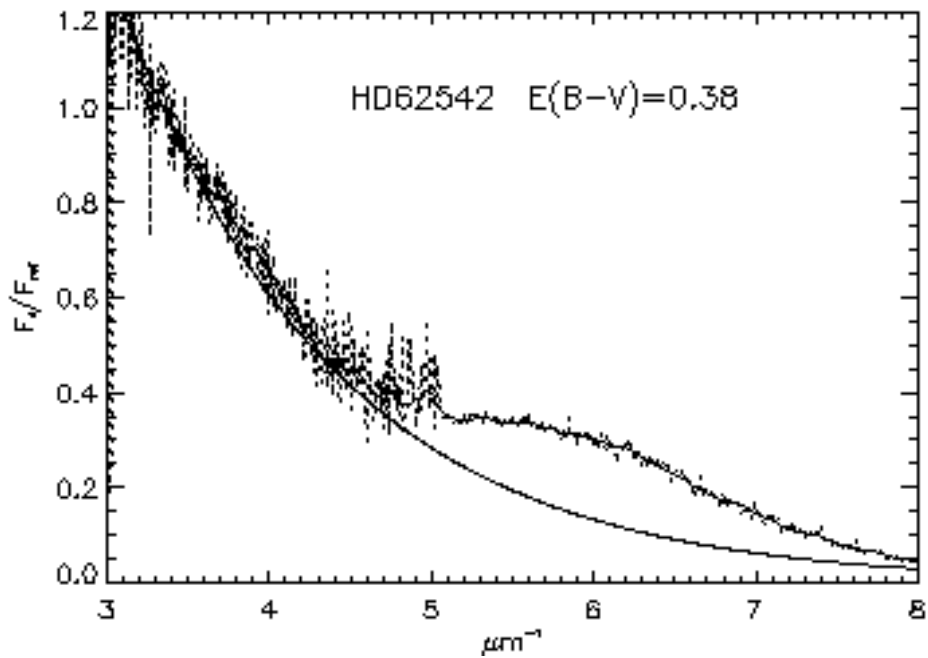

Figure 2 : Le spectre de HD62542 divisé par le spectre de l'étoile non-rougie de référence HD32630. L'extinction visible correspond à l'exponentielle et s'étend jusque dans la région du bump. Sur la queue de cette exponentielle se greffe un "excés" de lumière dû à l'apport de lumière diffusée.